\documentstyle[preprint,aps]{revtex}
\begin{document}
\draft
\preprint{\vbox{To be published in Jour. Math. Phys. 
\hfill  \hfill RCNP-Th00046 }}
\tolerance = 10000
\hfuzz=5pt

\title{Three-quark exchange operators, crossing matrices and Fierz 
transformations in SU(2) and SU(3)}
\author{V. Dmitra\v sinovi\' c }     
\address{Research Center for Nuclear Physics, Osaka University,\\
Mihogaoka 10 - 1, Ibaraki, Osaka 567-0047, Japan  }
 


\date{\today}
\maketitle
\begin{abstract}
We give explicit expressions for the three-quark exchange operators,
crossing matrices and Fierz transforms for the SU(2) and SU(3) groups.
We identify the invariant terms in these operators and express them 
in terms of Casimir operators.   
\end{abstract}
\widetext

\section{Introduction}

Dirac \cite{Dirac} was the first to express the two-particle spin-exchange 
operator 
\begin{eqnarray}
{\rm P}_{12} &=& {1 \over 2} + {1 \over 2}
\mbox{\boldmath$\tau$}_{1} \cdot \mbox{\boldmath$\tau$}_{2}
\label{e:12}\
\end{eqnarray}
in terms of Pauli matrices $\mbox{\boldmath$\tau$}_{1,2}$ of the two 
particles, where $\mbox{\boldmath$\tau$} \cdot \mbox{\boldmath$\tau$}
= \sum_{a = 1}^{3} \tau^{a}  \tau^{a}$.
This, and the following result
\begin{eqnarray}
{\rm P}_{12}\mbox{\boldmath$\tau$}_{1} \cdot \mbox{\boldmath$\tau$}_{2} 
&=& {3 \over 2} - {1 \over 2}
\mbox{\boldmath$\tau$}_{1} \cdot \mbox{\boldmath$\tau$}_{2}
\label{e:12'}\
\end{eqnarray}
are equivalent to a ``Fierz reordering
formula'' for the quartic field interaction, or to the SU(2) crossing 
matrix 
\begin{eqnarray}
{\bf C} &=& {1 \over 2}
\left(\begin{array}{cc}
1 &~1 \\
3 &-1 \
\end{array}\right)~.
\label{e:c2su2}
\end{eqnarray}   
for two-body processes. 
The same trick has been extended to the
SU(3) Lie algebra 
\begin{mathletters}
\begin{eqnarray}
{\rm P}_{12} &=& {1 \over 3} + {1 \over 2}
\mbox{\boldmath$\lambda$}_{1} \cdot \mbox{\boldmath$\lambda$}_{2}
\label{e:12su3} \\
{\rm P}_{12}\mbox{\boldmath$\lambda$}_{1} \cdot \mbox{\boldmath$\lambda$}_{2} 
&=& {16 \over 9} - {1 \over 3}
\mbox{\boldmath$\lambda$}_{1} \cdot \mbox{\boldmath$\lambda$}_{2}
\label{e:12'su3} \
\end{eqnarray}
\end{mathletters}
again only for two particles \cite{Okun}, with the resulting 
crossing matrix (or the equivalent Fierz reordering formulas for 
bilinear products of Gell-Mann matrices) being
\begin{eqnarray}
{\bf C} &=& 
\left(\begin{array}{cc}
~{1 \over 3} &~{1 \over 2} \\
{16 \over 9} &-{1 \over 3} \
\end{array}\right)~.
\label{e:c2su3}
\end{eqnarray}   
Here the lower index 
indicates the number of the quark, $\lambda^{a}$ are the Gell-Mann 
matrices, and $f^{abc}$, $d^{abc}$ are the usual SU(3) structure constants.

In the meantime a need has arisen for tri-linear Fierz 
formulas/crossing relations in connection with applications of the 
three-flavour `t Hooft interaction \cite{svz80}. Such relations don't 
seem to be extant in the literature \cite{BB63,Pais66,Carruth}.
In this note we present the corresponding three-body exchange operators 
for quarks (particles in the fundamental representation of SU(2) and/or 
SU(3)), as well as 
the equivalent Fierz reordering formulas for the sextic field 
interaction.

\section{Three-quark exchange operators} 

\subsection{The SU(2) algebra}

\subsubsection{Three-body exchange operators}

Using the 
symmetric group ${\cal S}_{3}$ it is 
straightforward, if tedious, to derive
the SU(2) version of the three-quark/spin exchange operator
\begin{mathletters}
\begin{eqnarray}
{\rm P}_{123} &=& {\rm P}_{23} {\rm P}_{12}
\nonumber \\
&=& {1 \over 4} + {1 \over 4}
\sum_{i < j}^{3} \mbox{\boldmath$\tau$}_{i} \cdot 
\mbox{\boldmath$\tau$}_{j}
 + {i \over 4} \varepsilon^{abc} \mbox{\boldmath$\tau$}_{1}^{a} 
\mbox{\boldmath$\tau$}_{2}^{b} \mbox{\boldmath$\tau$}_{3}^{c} 
\label{e:123su2}\\
{\rm P}_{132} &=& {\rm P}_{123}^{2}
\nonumber \\
&=& {1 \over 4} + {1 \over 4}
\sum_{i < j}^{3} \mbox{\boldmath$\tau$}_{i} \cdot 
\mbox{\boldmath$\tau$}_{j}
-{i \over 4} \varepsilon^{abc} \mbox{\boldmath$\tau$}_{1}^{a} 
\mbox{\boldmath$\tau$}_{2}^{b} \mbox{\boldmath$\tau$}_{3}^{c} 
\label{e:132su2}\
\end{eqnarray}
\end{mathletters}
Similar results 
are
\begin{mathletters}
\begin{eqnarray}
{\rm P}_{123} \sum_{i < j}^{3} \mbox{\boldmath$\tau$}_{i} \cdot 
\mbox{\boldmath$\tau$}_{j} 
&=& {1 \over 2} \left(9 + 
\sum_{i < j}^{3} \mbox{\boldmath$\tau$}_{i} \cdot 
\mbox{\boldmath$\tau$}_{j}
 - 3 i \varepsilon^{abc} \mbox{\boldmath$\tau$}_{1}^{a} 
\mbox{\boldmath$\tau$}_{2}^{b} \mbox{\boldmath$\tau$}_{3}^{c}  \right)
\label{e:123ssu2}\\
{\rm P}_{132} \sum_{i < j}^{3} \mbox{\boldmath$\tau$}_{i} \cdot 
\mbox{\boldmath$\tau$}_{j}
&=& {1 \over 4} \left(9 + 
\sum_{i < j}^{3} \mbox{\boldmath$\tau$}_{i} \cdot 
\mbox{\boldmath$\tau$}_{j}
 + 3 i \varepsilon^{abc} \mbox{\boldmath$\tau$}_{1}^{a} 
\mbox{\boldmath$\tau$}_{2}^{b} \mbox{\boldmath$\tau$}_{3}^{c}  \right)
\label{e:132ssu2}\
\end{eqnarray}
\end{mathletters}
as well as
\begin{mathletters}
\begin{eqnarray}
i {\rm P}_{123} \varepsilon^{abc} \mbox{\boldmath$\tau$}_{1}^{a} 
\mbox{\boldmath$\tau$}_{2}^{b} \mbox{\boldmath$\tau$}_{3}^{c}
&=& {1 \over 2} \left(- 3 +  
\sum_{i < j}^{3} \mbox{\boldmath$\tau$}_{i} \cdot 
\mbox{\boldmath$\tau$}_{j}
 - i \varepsilon^{abc} \mbox{\boldmath$\tau$}_{1}^{a} 
\mbox{\boldmath$\tau$}_{2}^{b} \mbox{\boldmath$\tau$}_{3}^{c}  \right)
\label{e:123esu2}\\
i {\rm P}_{132} \varepsilon^{abc} \mbox{\boldmath$\tau$}_{1}^{a} 
\mbox{\boldmath$\tau$}_{2}^{b} \mbox{\boldmath$\tau$}_{3}^{c}
&=& {1 \over 2} \left(3 - 
\sum_{i < j}^{3} \mbox{\boldmath$\tau$}_{i} \cdot 
\mbox{\boldmath$\tau$}_{j}
 - i \varepsilon^{abc} \mbox{\boldmath$\tau$}_{1}^{a} 
\mbox{\boldmath$\tau$}_{2}^{b} \mbox{\boldmath$\tau$}_{3}^{c}  \right)
\label{e:132esu2}\
\end{eqnarray}
\end{mathletters}

\subsubsection{Crossing matrix}
These results are summarized by the crossing matrices
\begin{eqnarray}
{\bf C} &=& {1 \over 4}
\left(\begin{array}{ccc}
~~1 & ~~1 &~~1 \\
~~9 & ~~1 &-3 \\
 -6 & ~~2 &-2 \
\end{array}\right)~,
\label{e:c3su2}
\end{eqnarray}  
for ${\rm P}_{123}$, and
\begin{eqnarray}
{\bf C}^{2} &=& {1 \over 4}
\left(\begin{array}{ccc}
~1 & ~~1 &-1 \\
 ~9 & ~~1 &~~3 \\
 6 & -2 &-2 \
\end{array}\right)~,
\label{e:c3'su2}
\end{eqnarray} 
for ${\rm P}_{132}$.
A valuable check is the constraint
${\bf C}^{3} = {\it 1}$.

\subsubsection{Three-body Fierz identities}

\begin{mathletters}
\begin{eqnarray}
{\delta}_{\alpha \delta} {\delta}_{\gamma \rho}
{\delta}_{\sigma \beta} &=& 
{1 \over 4}\Bigg({\delta}_{\alpha \beta} {\delta}_{\gamma \delta}
{\delta}_{\sigma \rho} + 
\delta_{\alpha \beta} \mbox{\boldmath$\tau$}_{\gamma \delta} \cdot 
\mbox{\boldmath$\tau$}_{\sigma \rho} +
\delta_{\gamma \delta} \mbox{\boldmath$\tau$}_{\alpha \beta} \cdot 
\mbox{\boldmath$\tau$}_{\sigma \rho} 
\nonumber \\
&+& 
\delta_{\sigma \rho} \mbox{\boldmath$\tau$}_{\gamma \delta} \cdot 
\mbox{\boldmath$\tau$}_{\alpha \beta} 
+ i \varepsilon^{abc} \mbox{\boldmath$\tau$}_{\alpha \beta}^{a} 
\mbox{\boldmath$\tau$}_{\gamma \delta}^{b} 
\mbox{\boldmath$\tau$}_{\sigma \rho}^{c} \Bigg)
\label{e:3bsu2}\\
\delta_{\alpha \rho} {\delta}_{\gamma \beta}
{\delta}_{\sigma \delta} &=& 
{1 \over 4}\Bigg({\delta}_{\alpha \beta} {\delta}_{\gamma \delta}
{\delta}_{\sigma \rho} + 
\delta_{\alpha \beta} \mbox{\boldmath$\tau$}_{\gamma \delta} \cdot 
\mbox{\boldmath$\tau$}_{\sigma \rho} +
\delta_{\gamma \delta} \mbox{\boldmath$\tau$}_{\alpha \beta} \cdot 
\mbox{\boldmath$\tau$}_{\sigma \rho} 
\nonumber \\
&+& 
\delta_{\sigma \rho} \mbox{\boldmath$\tau$}_{\gamma \delta} \cdot 
\mbox{\boldmath$\tau$}_{\alpha \beta} 
- i \varepsilon^{abc} \mbox{\boldmath$\tau$}_{\alpha \beta}^{a} 
\mbox{\boldmath$\tau$}_{\gamma \delta}^{b} 
\mbox{\boldmath$\tau$}_{\sigma \rho}^{c} \Bigg)
\label{e:3b'su2}\
\end{eqnarray}
\end{mathletters}

\begin{mathletters}
\begin{eqnarray}
&&
{\delta}_{\alpha \delta} \mbox{\boldmath$\tau$}_{\gamma \rho} \cdot 
\mbox{\boldmath$\tau$}_{\sigma \beta} + 
{\delta}_{\gamma \rho} \mbox{\boldmath$\tau$}_{\alpha \delta} \cdot 
\mbox{\boldmath$\tau$}_{\sigma \beta} +
{\delta}_{\sigma \beta} \mbox{\boldmath$\tau$}_{\gamma \rho} \cdot 
\mbox{\boldmath$\tau$}_{\alpha \delta}
= 
{1 \over 4}\Bigg(9 
{\delta}_{\alpha \beta} {\delta}_{\gamma \delta} {\delta}_{\sigma \rho} 
\nonumber \\
&+& 
\delta_{\alpha \beta} \mbox{\boldmath$\tau$}_{\gamma \delta} \cdot 
\mbox{\boldmath$\tau$}_{\sigma \rho} +
\delta_{\gamma \delta} \mbox{\boldmath$\tau$}_{\alpha \beta} \cdot 
\mbox{\boldmath$\tau$}_{\sigma \rho} +
\delta_{\sigma \rho} \mbox{\boldmath$\tau$}_{\gamma \delta} \cdot 
\mbox{\boldmath$\tau$}_{\alpha \beta} 
 - 3 i \varepsilon^{abc} \mbox{\boldmath$\tau$}_{\alpha \beta}^{a} 
\mbox{\boldmath$\tau$}_{\gamma \delta}^{b} 
\mbox{\boldmath$\tau$}_{\sigma \rho}^{c} \Bigg)
\label{e:3bssu2}\\
&&
{\delta}_{\alpha \rho} \mbox{\boldmath$\tau$}_{\gamma \beta} \cdot 
\mbox{\boldmath$\tau$}_{\sigma \delta} +
{\delta}_{\gamma \beta} 
\mbox{\boldmath$\tau$}_{\alpha \rho} \cdot 
\mbox{\boldmath$\tau$}_{\sigma \delta} +
{\delta}_{\sigma \beta} \mbox{\boldmath$\tau$}_{\gamma \rho} \cdot 
\mbox{\boldmath$\tau$}_{\alpha \delta}
= 
{1 \over 4}\Bigg(9
{\delta}_{\alpha \beta} {\delta}_{\gamma \delta}{\delta}_{\sigma \rho} 
\nonumber \\
&+& 
\delta_{\alpha \beta} \mbox{\boldmath$\tau$}_{\gamma \delta} \cdot 
\mbox{\boldmath$\tau$}_{\sigma \rho} +
\delta_{\gamma \delta} \mbox{\boldmath$\tau$}_{\alpha \beta} \cdot 
\mbox{\boldmath$\tau$}_{\sigma \rho} +
\delta_{\sigma \rho} \mbox{\boldmath$\tau$}_{\gamma \delta} \cdot 
\mbox{\boldmath$\tau$}_{\alpha \beta} 
+ 3 i \varepsilon^{abc} \mbox{\boldmath$\tau$}_{\alpha \beta}^{a} 
\mbox{\boldmath$\tau$}_{\gamma \delta}^{b} 
\mbox{\boldmath$\tau$}_{\sigma \rho}^{c} \Bigg)
\label{e:3bs'su2}\
\end{eqnarray}
\end{mathletters}

\begin{mathletters}
\begin{eqnarray}
&&
i \varepsilon^{abc} \mbox{\boldmath$\tau$}_{\alpha \delta}^{a} 
\mbox{\boldmath$\tau$}_{\gamma \rho}^{b} 
\mbox{\boldmath$\tau$}_{\sigma \beta}^{c}
= 
{1 \over 2}\Bigg( - 3 
{\delta}_{\alpha \beta} {\delta}_{\gamma \delta} {\delta}_{\sigma \rho} 
\nonumber \\
&+& 
\delta_{\alpha \beta} \mbox{\boldmath$\tau$}_{\gamma \delta} \cdot 
\mbox{\boldmath$\tau$}_{\sigma \rho} +
\delta_{\gamma \delta} \mbox{\boldmath$\tau$}_{\alpha \beta} \cdot 
\mbox{\boldmath$\tau$}_{\sigma \rho} +
\delta_{\sigma \rho} \mbox{\boldmath$\tau$}_{\gamma \delta} \cdot 
\mbox{\boldmath$\tau$}_{\alpha \beta} 
 - i \varepsilon^{abc} \mbox{\boldmath$\tau$}_{\alpha \beta}^{a} 
\mbox{\boldmath$\tau$}_{\gamma \delta}^{b} 
\mbox{\boldmath$\tau$}_{\sigma \rho}^{c} \Bigg)
\label{e:3bfsu2}\\
&&
i \varepsilon^{abc} \mbox{\boldmath$\tau$}_{\alpha \rho}^{a} 
\mbox{\boldmath$\tau$}_{\gamma \beta}^{b} 
\mbox{\boldmath$\tau$}_{\sigma \delta}^{c}
= 
{1 \over 2}\Bigg(3
{\delta}_{\alpha \beta} {\delta}_{\gamma \delta}{\delta}_{\sigma \rho} 
\nonumber \\
&-& 
\left(\delta_{\alpha \beta} \mbox{\boldmath$\tau$}_{\gamma \delta} \cdot 
\mbox{\boldmath$\tau$}_{\sigma \rho} +
\delta_{\gamma \delta} \mbox{\boldmath$\tau$}_{\alpha \beta} \cdot 
\mbox{\boldmath$\tau$}_{\sigma \rho} +
\delta_{\sigma \rho} \mbox{\boldmath$\tau$}_{\gamma \delta} \cdot 
\mbox{\boldmath$\tau$}_{\alpha \beta} \right)
- i f^{abc} \mbox{\boldmath$\tau$}_{\alpha \beta}^{a} 
\mbox{\boldmath$\tau$}_{\gamma \delta}^{b} 
\mbox{\boldmath$\tau$}_{\sigma \rho}^{c} \Bigg)
\label{e:3bf'su2}\
\end{eqnarray}
\end{mathletters}

\subsection{The SU(3) algebra}
\subsubsection{Three-body exchange operators}

Similarly, we have
\begin{mathletters}
\begin{eqnarray}
{\rm P}_{123} &=& {1 \over 9} + {1 \over 6}
\sum_{i < j}^{3} \mbox{\boldmath$\lambda$}_{i} \cdot 
\mbox{\boldmath$\lambda$}_{j}
 + {1 \over 4} d^{abc} \mbox{\boldmath$\lambda$}_{1}^{a} 
\mbox{\boldmath$\lambda$}_{2}^{b} \mbox{\boldmath$\lambda$}_{3}^{c} 
 + {i \over 4} f^{abc} \mbox{\boldmath$\lambda$}_{1}^{a} 
\mbox{\boldmath$\lambda$}_{2}^{b} \mbox{\boldmath$\lambda$}_{3}^{c} 
\label{e:123su3}\\
{\rm P}_{132} &=& {1 \over 9} + {1 \over 6}
\sum_{i < j}^{3} \mbox{\boldmath$\lambda$}_{i} \cdot 
\mbox{\boldmath$\lambda$}_{j}
 + {1 \over 4} d^{abc} \mbox{\boldmath$\lambda$}_{1}^{a} 
\mbox{\boldmath$\lambda$}_{2}^{b} \mbox{\boldmath$\lambda$}_{3}^{c} 
 - {1 \over 4} i f^{abc} \mbox{\boldmath$\lambda$}_{1}^{a} 
\mbox{\boldmath$\lambda$}_{2}^{b} \mbox{\boldmath$\lambda$}_{3}^{c} 
\label{e:132su3}\
\end{eqnarray}
\end{mathletters}
as well as similar relations for the operators
\begin{mathletters}
\begin{eqnarray}
{\rm P}_{123} \sum_{i < j}^{3} \mbox{\boldmath$\lambda$}_{i} \cdot 
\mbox{\boldmath$\lambda$}_{j}
&=& {1 \over 9} + {1 \over 6}
\sum_{i < j}^{3} \mbox{\boldmath$\lambda$}_{i} \cdot 
\mbox{\boldmath$\lambda$}_{j}
 + {1 \over 4} d^{abc} \mbox{\boldmath$\lambda$}_{1}^{a} 
\mbox{\boldmath$\lambda$}_{2}^{b} \mbox{\boldmath$\lambda$}_{3}^{c} 
 + {i \over 4} f^{abc} \mbox{\boldmath$\lambda$}_{1}^{a} 
\mbox{\boldmath$\lambda$}_{2}^{b} \mbox{\boldmath$\lambda$}_{3}^{c} 
\label{e:123ssu3}\\
{\rm P}_{132} \sum_{i < j}^{3} \mbox{\boldmath$\lambda$}_{i} \cdot 
\mbox{\boldmath$\lambda$}_{j}
&=& {1 \over 9} + {1 \over 6}
\sum_{i < j}^{3} \mbox{\boldmath$\lambda$}_{i} \cdot 
\mbox{\boldmath$\lambda$}_{j}
 + {1 \over 4} d^{abc} \mbox{\boldmath$\lambda$}_{1}^{a} 
\mbox{\boldmath$\lambda$}_{2}^{b} \mbox{\boldmath$\lambda$}_{3}^{c} 
 - {1 \over 4} i f^{abc} \mbox{\boldmath$\lambda$}_{1}^{a} 
\mbox{\boldmath$\lambda$}_{2}^{b} \mbox{\boldmath$\lambda$}_{3}^{c} 
\label{e:132ssu3}\
\end{eqnarray}
\end{mathletters}
and
\begin{mathletters}
\begin{eqnarray}
{\rm P}_{123} d^{abc} \mbox{\boldmath$\lambda$}_{1}^{a} 
\mbox{\boldmath$\lambda$}_{2}^{b} \mbox{\boldmath$\lambda$}_{3}^{c} 
&=& 
{80 \over 81} - {5 \over 27}
\sum_{i < j}^{3} \mbox{\boldmath$\lambda$}_{i} \cdot 
\mbox{\boldmath$\lambda$}_{j}
 + {13 \over 18} d^{abc} \mbox{\boldmath$\lambda$}_{1}^{a} 
\mbox{\boldmath$\lambda$}_{2}^{b} \mbox{\boldmath$\lambda$}_{3}^{c} 
 -{5 \over 18} i f^{abc} \mbox{\boldmath$\lambda$}_{1}^{a} 
\mbox{\boldmath$\lambda$}_{2}^{b} \mbox{\boldmath$\lambda$}_{3}^{c} 
\label{e:123dsu3}\\
{\rm P}_{132} d^{abc} \mbox{\boldmath$\lambda$}_{1}^{a} 
\mbox{\boldmath$\lambda$}_{2}^{b} \mbox{\boldmath$\lambda$}_{3}^{c} 
&=& 
{80 \over 81} - {5 \over 27}
\sum_{i < j}^{3} \mbox{\boldmath$\lambda$}_{i} \cdot 
\mbox{\boldmath$\lambda$}_{j}
 + {13 \over 18} d^{abc} \mbox{\boldmath$\lambda$}_{1}^{a} 
\mbox{\boldmath$\lambda$}_{2}^{b} \mbox{\boldmath$\lambda$}_{3}^{c} 
+ {5 \over 18} i f^{abc} \mbox{\boldmath$\lambda$}_{1}^{a} 
\mbox{\boldmath$\lambda$}_{2}^{b} \mbox{\boldmath$\lambda$}_{3}^{c} 
\label{e:132dsu3}\\
{\rm P}_{123} i f^{abc} \mbox{\boldmath$\lambda$}_{1}^{a} 
\mbox{\boldmath$\lambda$}_{2}^{b} \mbox{\boldmath$\lambda$}_{3}^{c} 
&=& 
-{16 \over 9} + {1 \over 3}
\sum_{i < j}^{3} \mbox{\boldmath$\lambda$}_{i} \cdot 
\mbox{\boldmath$\lambda$}_{j}
 + {1 \over 2} d^{abc} \mbox{\boldmath$\lambda$}_{1}^{a} 
\mbox{\boldmath$\lambda$}_{2}^{b} \mbox{\boldmath$\lambda$}_{3}^{c} 
- {1 \over 2} i f^{abc} \mbox{\boldmath$\lambda$}_{1}^{a} 
\mbox{\boldmath$\lambda$}_{2}^{b} \mbox{\boldmath$\lambda$}_{3}^{c} 
\label{e:123fsu3}\\
{\rm P}_{132} i f^{abc} \mbox{\boldmath$\lambda$}_{1}^{a} 
\mbox{\boldmath$\lambda$}_{2}^{b} \mbox{\boldmath$\lambda$}_{3}^{c} 
&=& 
~~ {16 \over 9} ~ - {1 \over 3}
\sum_{i < j}^{3} \mbox{\boldmath$\lambda$}_{i} \cdot 
\mbox{\boldmath$\lambda$}_{j}
 - {1 \over 2} d^{abc} \mbox{\boldmath$\lambda$}_{1}^{a} 
\mbox{\boldmath$\lambda$}_{2}^{b} \mbox{\boldmath$\lambda$}_{3}^{c} 
 - {1 \over 2} i f^{abc} \mbox{\boldmath$\lambda$}_{1}^{a} 
\mbox{\boldmath$\lambda$}_{2}^{b} \mbox{\boldmath$\lambda$}_{3}^{c} 
\label{e:132fsu3}\
\end{eqnarray}
\end{mathletters}

\subsubsection{Crossing matrix}

This leads to the following first cyclic permutation three-quark 
crossing matrix
\begin{eqnarray}
{\bf C} &=& 
\left(\begin{array}{cccccccc}
~{1 \over 9} && {1 \over 6} &&~{1 \over 4} &&~{1 \over 4}\\
~{16 \over 9} && {2 \over 3} &&-{1 \over 2} &&~{1 \over 2}\\
~{80 \over 81} && -{5 \over 27} &&~{13 \over 18} &&-{5 \over 18}\\
-{16 \over 9} && ~{1 \over 3} &&~{1 \over 2} &&-{1 \over 2}\
\end{array}\right)~.
\label{e:c3su3}
\end{eqnarray}   
Similarly, for the second cyclic permutation we find
\begin{eqnarray}
{\bf C}^{2} &=& 
\left(\begin{array}{cccccccc}
~{1 \over 9} && {1 \over 6} &&~{1 \over 4} &&-{1 \over 4}\\
~{16 \over 9} && {2 \over 3} &&-{1 \over 2} &&-{1 \over 2}\\
~{80 \over 81} && -{5 \over 27} &&~{13 \over 18} &&~{5 \over 18}\\
~{16 \over 9} && -{1 \over 3} &&-{1 \over 2} &&-{1 \over 2}\
\end{array}\right)~,
\label{e:c3'su3}
\end{eqnarray}   
and, of course satisfying ${\bf C}^{3} = {\it 1}$. 

\subsubsection{Three-body Fierz identities}

\begin{mathletters}
\begin{eqnarray}
{\delta}_{\alpha \delta} {\delta}_{\gamma \rho}
{\delta}_{\sigma \beta} &=& 
{1 \over 9}{\delta}_{\alpha \beta} {\delta}_{\gamma \delta}
{\delta}_{\sigma \rho} + 
{1 \over 6}\left(
\delta_{\alpha \beta} \mbox{\boldmath$\lambda$}_{\gamma \delta} \cdot 
\mbox{\boldmath$\lambda$}_{\sigma \rho} +
\delta_{\gamma \delta} \mbox{\boldmath$\lambda$}_{\alpha \beta} \cdot 
\mbox{\boldmath$\lambda$}_{\sigma \rho} +
\delta_{\sigma \rho} \mbox{\boldmath$\lambda$}_{\gamma \delta} \cdot 
\mbox{\boldmath$\lambda$}_{\alpha \beta} \right)
\nonumber \\
&+& 
{1 \over 4} d^{abc} \mbox{\boldmath$\lambda$}_{\alpha \beta}^{a} 
\mbox{\boldmath$\lambda$}_{\gamma \delta}^{b} 
\mbox{\boldmath$\lambda$}_{\sigma \rho}^{c}
+ {1 \over 4} i f^{abc} \mbox{\boldmath$\lambda$}_{\alpha \beta}^{a} 
\mbox{\boldmath$\lambda$}_{\gamma \delta}^{b} 
\mbox{\boldmath$\lambda$}_{\sigma \rho}^{c}
\label{e:3bsu3}\\
\delta_{\alpha \rho} {\delta}_{\gamma \beta}
{\delta}_{\sigma \delta} &=& 
{1 \over 9}{\delta}_{\alpha \beta} {\delta}_{\gamma \delta}
{\delta}_{\sigma \rho} + 
{1 \over 6}\left(
\delta_{\alpha \beta} \mbox{\boldmath$\lambda$}_{\gamma \delta} \cdot 
\mbox{\boldmath$\lambda$}_{\sigma \rho} +
\delta_{\gamma \delta} \mbox{\boldmath$\lambda$}_{\alpha \beta} \cdot 
\mbox{\boldmath$\lambda$}_{\sigma \rho} +
\delta_{\sigma \rho} \mbox{\boldmath$\lambda$}_{\gamma \delta} \cdot 
\mbox{\boldmath$\lambda$}_{\alpha \beta} \right)
\nonumber \\
&+& 
{1 \over 4} d^{abc} \mbox{\boldmath$\lambda$}_{\alpha \beta}^{a} 
\mbox{\boldmath$\lambda$}_{\gamma \delta}^{b} 
\mbox{\boldmath$\lambda$}_{\sigma \rho}^{c}
- {1 \over 4} i f^{abc} \mbox{\boldmath$\lambda$}_{\alpha \beta}^{a} 
\mbox{\boldmath$\lambda$}_{\gamma \delta}^{b} 
\mbox{\boldmath$\lambda$}_{\sigma \rho}^{c} 
\label{e:3b'su3}\
\end{eqnarray}
\end{mathletters}

\begin{mathletters}
\begin{eqnarray}
&&
{\delta}_{\alpha \delta} \mbox{\boldmath$\lambda$}_{\gamma \rho} \cdot 
\mbox{\boldmath$\lambda$}_{\sigma \beta} + 
{\delta}_{\gamma \rho} \mbox{\boldmath$\lambda$}_{\alpha \delta} \cdot 
\mbox{\boldmath$\lambda$}_{\sigma \beta} +
{\delta}_{\sigma \beta} \mbox{\boldmath$\lambda$}_{\gamma \rho} \cdot 
\mbox{\boldmath$\lambda$}_{\alpha \delta}
\nonumber \\
&=& 
{2 \over 3}
\left(\delta_{\alpha \beta} \mbox{\boldmath$\lambda$}_{\gamma \delta} \cdot 
\mbox{\boldmath$\lambda$}_{\sigma \rho} +
\delta_{\gamma \delta} \mbox{\boldmath$\lambda$}_{\alpha \beta} \cdot 
\mbox{\boldmath$\lambda$}_{\sigma \rho} +
\delta_{\sigma \rho} \mbox{\boldmath$\lambda$}_{\gamma \delta} \cdot 
\mbox{\boldmath$\lambda$}_{\alpha \beta} \right)
\nonumber \\
&+& 
{16 \over 9} 
{\delta}_{\alpha \beta} {\delta}_{\gamma \delta} {\delta}_{\sigma \rho} 
- {1 \over 2} d^{abc} \mbox{\boldmath$\lambda$}_{\alpha \beta}^{a} 
\mbox{\boldmath$\lambda$}_{\gamma \delta}^{b} 
\mbox{\boldmath$\lambda$}_{\sigma \rho}^{c}
 + {1 \over 2} i f^{abc} \mbox{\boldmath$\lambda$}_{\alpha \beta}^{a} 
\mbox{\boldmath$\lambda$}_{\gamma \delta}^{b} 
\mbox{\boldmath$\lambda$}_{\sigma \rho}^{c}
\label{e:3bssu3}\\
&&
{\delta}_{\alpha \rho} \mbox{\boldmath$\lambda$}_{\gamma \beta} \cdot 
\mbox{\boldmath$\lambda$}_{\sigma \delta} +
{\delta}_{\gamma \beta} 
\mbox{\boldmath$\lambda$}_{\alpha \rho} \cdot 
\mbox{\boldmath$\lambda$}_{\sigma \delta} +
{\delta}_{\sigma \beta} \mbox{\boldmath$\lambda$}_{\gamma \rho} \cdot 
\mbox{\boldmath$\lambda$}_{\alpha \delta}
\nonumber \\
&=& 
{2 \over 3}
\left(\delta_{\alpha \beta} \mbox{\boldmath$\lambda$}_{\gamma \delta} \cdot 
\mbox{\boldmath$\lambda$}_{\sigma \rho} +
\delta_{\gamma \delta} \mbox{\boldmath$\lambda$}_{\alpha \beta} \cdot 
\mbox{\boldmath$\lambda$}_{\sigma \rho} +
\delta_{\sigma \rho} \mbox{\boldmath$\lambda$}_{\gamma \delta} \cdot 
\mbox{\boldmath$\lambda$}_{\alpha \beta} \right)
\nonumber \\
&+& 
{16 \over 9} 
{\delta}_{\alpha \beta} {\delta}_{\gamma \delta} {\delta}_{\sigma \rho} 
- {1 \over 2} d^{abc} \mbox{\boldmath$\lambda$}_{\alpha \beta}^{a} 
\mbox{\boldmath$\lambda$}_{\gamma \delta}^{b} 
\mbox{\boldmath$\lambda$}_{\sigma \rho}^{c}
- {1 \over 2} i f^{abc} \mbox{\boldmath$\lambda$}_{\alpha \beta}^{a} 
\mbox{\boldmath$\lambda$}_{\gamma \delta}^{b} 
\mbox{\boldmath$\lambda$}_{\sigma \rho}^{c}
\label{e:3bs'su3}\
\end{eqnarray}
\end{mathletters}

\begin{mathletters}
\begin{eqnarray}
d^{abc} \mbox{\boldmath$\lambda$}_{\alpha \delta}^{a} 
\mbox{\boldmath$\lambda$}_{\gamma \rho}^{b} 
\mbox{\boldmath$\lambda$}_{\sigma \beta}^{c}
&=& 
-{5 \over 27}
\left(\delta_{\alpha \beta} \mbox{\boldmath$\lambda$}_{\gamma \delta} \cdot 
\mbox{\boldmath$\lambda$}_{\sigma \rho} +
\delta_{\gamma \delta} \mbox{\boldmath$\lambda$}_{\alpha \beta} \cdot 
\mbox{\boldmath$\lambda$}_{\sigma \rho} +
\delta_{\sigma \rho} \mbox{\boldmath$\lambda$}_{\gamma \delta} \cdot 
\mbox{\boldmath$\lambda$}_{\alpha \beta} \right)
\nonumber \\
&+& 
{80 \over 81} 
{\delta}_{\alpha \beta} {\delta}_{\gamma \delta} {\delta}_{\sigma \rho} 
+ {13 \over 18} d^{abc} \mbox{\boldmath$\lambda$}_{\alpha \beta}^{a} 
\mbox{\boldmath$\lambda$}_{\gamma \delta}^{b} 
\mbox{\boldmath$\lambda$}_{\sigma \rho}^{c}
- {5 \over 18} i f^{abc} \mbox{\boldmath$\lambda$}_{\alpha \beta}^{a} 
\mbox{\boldmath$\lambda$}_{\gamma \delta}^{b} 
\mbox{\boldmath$\lambda$}_{\sigma \rho}^{c}
\label{e:3bdsu3}\\
d^{abc} \mbox{\boldmath$\lambda$}_{\alpha \rho}^{a} 
\mbox{\boldmath$\lambda$}_{\gamma \beta}^{b} 
\mbox{\boldmath$\lambda$}_{\sigma \delta}^{c}
&=& 
-{5 \over 27}
\left(\delta_{\alpha \beta} \mbox{\boldmath$\lambda$}_{\gamma \delta} \cdot 
\mbox{\boldmath$\lambda$}_{\sigma \rho} +
\delta_{\gamma \delta} \mbox{\boldmath$\lambda$}_{\alpha \beta} \cdot 
\mbox{\boldmath$\lambda$}_{\sigma \rho} +
\delta_{\sigma \rho} \mbox{\boldmath$\lambda$}_{\gamma \delta} \cdot 
\mbox{\boldmath$\lambda$}_{\alpha \beta} \right)
\nonumber \\
&+& 
{80 \over 81} 
{\delta}_{\alpha \beta} {\delta}_{\gamma \delta} {\delta}_{\sigma \rho} 
+ {13 \over 18} d^{abc} \mbox{\boldmath$\lambda$}_{\alpha \beta}^{a} 
\mbox{\boldmath$\lambda$}_{\gamma \delta}^{b} 
\mbox{\boldmath$\lambda$}_{\sigma \rho}^{c}
+ {5 \over 18} i f^{abc} \mbox{\boldmath$\lambda$}_{\alpha \beta}^{a} 
\mbox{\boldmath$\lambda$}_{\gamma \delta}^{b} 
\mbox{\boldmath$\lambda$}_{\sigma \rho}^{c}
\label{e:3bd'su3}\
\end{eqnarray}
\end{mathletters}

\begin{mathletters}
\begin{eqnarray}
i f^{abc} \mbox{\boldmath$\lambda$}_{\alpha \delta}^{a} 
\mbox{\boldmath$\lambda$}_{\gamma \rho}^{b} 
\mbox{\boldmath$\lambda$}_{\sigma \beta}^{c}
&=& 
{1 \over 3}
\left(\delta_{\alpha \beta} \mbox{\boldmath$\lambda$}_{\gamma \delta} \cdot 
\mbox{\boldmath$\lambda$}_{\sigma \rho} +
\delta_{\gamma \delta} \mbox{\boldmath$\lambda$}_{\alpha \beta} \cdot 
\mbox{\boldmath$\lambda$}_{\sigma \rho} +
\delta_{\sigma \rho} \mbox{\boldmath$\lambda$}_{\gamma \delta} \cdot 
\mbox{\boldmath$\lambda$}_{\alpha \beta} \right)
\nonumber \\
&-& 
{16 \over 9} 
{\delta}_{\alpha \beta} {\delta}_{\gamma \delta} {\delta}_{\sigma \rho} 
+ {1 \over 2} d^{abc} \mbox{\boldmath$\lambda$}_{\alpha \beta}^{a} 
\mbox{\boldmath$\lambda$}_{\gamma \delta}^{b} 
\mbox{\boldmath$\lambda$}_{\sigma \rho}^{c}
- {1 \over 2} i f^{abc} \mbox{\boldmath$\lambda$}_{\alpha \beta}^{a} 
\mbox{\boldmath$\lambda$}_{\gamma \delta}^{b} 
\mbox{\boldmath$\lambda$}_{\sigma \rho}^{c}
\label{e:3bfsu3}\\
i f^{abc} \mbox{\boldmath$\lambda$}_{\alpha \rho}^{a} 
\mbox{\boldmath$\lambda$}_{\gamma \beta}^{b} 
\mbox{\boldmath$\lambda$}_{\sigma \delta}^{c}
&=& 
-{1 \over 3}
\left(\delta_{\alpha \beta} \mbox{\boldmath$\lambda$}_{\gamma \delta} \cdot 
\mbox{\boldmath$\lambda$}_{\sigma \rho} +
\delta_{\gamma \delta} \mbox{\boldmath$\lambda$}_{\alpha \beta} \cdot 
\mbox{\boldmath$\lambda$}_{\sigma \rho} +
\delta_{\sigma \rho} \mbox{\boldmath$\lambda$}_{\gamma \delta} \cdot 
\mbox{\boldmath$\lambda$}_{\alpha \beta} \right)
\nonumber \\
&+& 
{16 \over 9} 
{\delta}_{\alpha \beta} {\delta}_{\gamma \delta} {\delta}_{\sigma \rho} 
- {1 \over 2} d^{abc} \mbox{\boldmath$\lambda$}_{\alpha \beta}^{a} 
\mbox{\boldmath$\lambda$}_{\gamma \delta}^{b} 
\mbox{\boldmath$\lambda$}_{\sigma \rho}^{c}
- {1 \over 2} i f^{abc} \mbox{\boldmath$\lambda$}_{\alpha \beta}^{a} 
\mbox{\boldmath$\lambda$}_{\gamma \delta}^{b} 
\mbox{\boldmath$\lambda$}_{\sigma \rho}^{c}
\label{e:3bf'su3}\
\end{eqnarray}
\end{mathletters}

\section{Comments}

Two of the three operators
$\sum_{i < j}^{3} \mbox{\boldmath$\lambda$}_{i} \cdot 
\mbox{\boldmath$\lambda$}_{j},
d^{abc} \mbox{\boldmath$\lambda$}_{1}^{a} 
\mbox{\boldmath$\lambda$}_{2}^{b} \mbox{\boldmath$\lambda$}_{3}^{c},
i f^{abc} \mbox{\boldmath$\lambda$}_{1}^{a} 
\mbox{\boldmath$\lambda$}_{2}^{b} \mbox{\boldmath$\lambda$}_{3}^{c} $
are SU(3) invariants,
i.e. they can be expressed in terms of the two Casimir operators of 
SU(3) as follows
\begin{mathletters}
\begin{eqnarray}
\sum_{i < j}^{3} \mbox{\boldmath$\lambda$}_{i} \cdot 
\mbox{\boldmath$\lambda$}_{j} &=& 
2 C^{(1)} - 4 \\
\label{e:sum1}
d^{abc} \mbox{\boldmath$\lambda$}_{1}^{a} 
\mbox{\boldmath$\lambda$}_{2}^{b} \mbox{\boldmath$\lambda$}_{3}^{c} 
&=& 
{4 \over 3} \left[C^{(2)} - 
{5 \over 2}C^{(1)} + {20 \over 3}\right]; \
\label{e:d1}
\end{eqnarray}
\end{mathletters}
where the two Casimir operators of SU(3) are defined as
$C^{(1)} = {\bf F}^{2}, C^{(2)} = d^{abc}{\bf F}^{a} {\bf F}^{b} {\bf F}^{c}$
and
${\bf F}^{a}$
are the SU(3) algebra generators. The third operator,
$i f^{abc} \mbox{\boldmath$\lambda$}_{1}^{a} 
\mbox{\boldmath$\lambda$}_{2}^{b} \mbox{\boldmath$\lambda$}_{3}^{c} $,
is a peculiar object: it is an SU(3) invariant, 
because it commutes with the SU(3) generators 
${\bf F}^{a} = {1 \over 2} \left(\mbox{\boldmath$\lambda$}_{1}^{a} + 
\mbox{\boldmath$\lambda$}_{2}^{a} + 
\mbox{\boldmath$\lambda$}_{3}^{a}\right)$
in the special case
when these generators are formed from three Gell-Mann matrices, 
but it is also an off-diagonal operator
[it annihilates the two three-quark SU(3) eigenstates with definite 
exchange symmetry properties, i.e. the {\bf 1} and {\bf 10}, 
and turns one {\bf 8} state into another] that cannot be expressed in 
terms of Casimir operators. 
This result does not violate the Casimir-v.d. Waerden
theorem relating the rank of the group to the number of independent 
invariant operators, as it is representation dependent. This example, 
however, points out the existence of such invariants which is not 
widely known.

\end{document}